\documentclass[iop]{emulateapj}

\def\spose#1{\hbox to 0pt{#1\hss}}
\def\approxlt{\mathrel{\spose{\lower 3pt\hbox{$\sim$}}
	\raise 2.0pt\hbox{$<$}}}
\def\approxgt{\mathrel{\spose{\lower 3pt\hbox{$\sim$}}
	\raise 2.0pt\hbox{$>$}}}
\def\approxpropto{\mathrel{\spose{\lower 3pt\hbox{$\sim$}}
	\raise 2.0pt\hbox{$\propto$}}}
\mathchardef\twiddle="2218

\def\multleft#1{\hbox to size{\vbox {\halign {\lft{##}\cr #1}}\hfill}\par}
\def\multright#1{\hbox to size{\vbox {\halign {\rt{##}\cr #1}}\hfill}\par}

\def\today{\ifcase\month\or January\or February\or March\or April\or May\or
      June\or July\or August\or September\or October\or November\or December\fi
      \space\number\day, \number\year}
\def\<{\thinspace}
\def\arcsec{{\rm\thinspace arcsec}}
\def\cm{{\rm\thinspace cm}}

\def\K{{\rm\thinspace K}}
\def\keV{{\rm\thinspace keV}}

\def\km{{\rm\thinspace km}}
\def\kpc{{\rm\thinspace kpc}}

\def\Mpc{{\rm\thinspace Mpc}}

\def\pc{{\rm\thinspace pc}}

\def\s{{\rm\thinspace s}}
\def\yr{{\rm\thinspace yr}}

\usepackage{graphicx}
\usepackage{txfonts}
\usepackage{natbib}
\usepackage{wrapfig}                  
\usepackage{amssymb}
\usepackage{longtable}
\usepackage{times}
\usepackage[usenames]{color}
\usepackage{psfig}
\usepackage{soul,color}
\usepackage{subfigure}
\newcommand*{\mysub}[2]{\ensuremath{#1_{\mathrm{#2}}}}
\newcommand*{\Omegam}{\mysub{\Omega}{m}}

\newcommand*{\Omegal}{\ensuremath{\Omega_{\Lambda}}}

\newcommand*{\msolar}{\mysub{M}{\odot}}
\newcommand*{\zsolar}{\mysub{Z}{\odot}}

\begin{document}

\title{A very deep Chandra observation of Abell 1795:\\ 
	 The Cold Front and and Cooling Wake
}
\author{Steven Ehlert\altaffilmark{1}, Michael McDonald\altaffilmark{1}, Laurence P.~David\altaffilmark{2}, Eric D.~Miller\altaffilmark{1}, \& Mark W.~Bautz\altaffilmark{1}}
\altaffiltext{*}{Email: sehlert@space.mit.edu}
\altaffiltext{1}{Kavli Institute for Astrophysics and Space Research, Massachusetts Institute of Technology, 77 Massachusetts Avenue,\\
Cambridge, MA 02139, USA;}
\altaffiltext{2}{Harvard-Smithsonian Astrophysical Observatory, 60 Garden Street, Cambridge, MA 02138, USA}

\def\cha{{\it Chandra}}
\def\suz{{\it Suzaku}}
\def\ae{{\small ACIS-EXTRACT}}
\def\xspec{{\small XSPEC}}
\def\marx{{\small MARX}}
\def\sub{{\it Subaru}}
\def\cfht{{\it CFHT}}
\def\mach{{\mathcal{M} }}
\def\rfive{\mysub{r}{500}}
\def\rosat{{\it ROSAT}}
\def\rtwo{ \mysub{r}{200}}
\def\xmm{{\it XMM-Newton}}
\def\cosm{{\it COSMOS}}

\def\sext{{\small SEXTRACTOR}}

\def\arcsec {\hbox{$^{\prime\prime}$}} 
\def\arcmin {\hbox{$^{\prime}$}} 
\def\MACS {MACS J1931.8-2634}

\def\arcsecf {\hbox{$.\!\!^{\prime\prime}$}} 
\def\arcminf {\hbox{$.\!\!^{\prime}$}}
\def\degree {\hbox{$.\!\!^{\circ}$}}
\def\wav{{\small WAVDETECT}}
\def\rad {\hbox{2\arcsec}}

\def\cdfs{{\it CDFS}}
\def\cdfn{{\it CDFN}}

\begin{abstract}
We present a new analysis of very deep \cha \ observations  of the galaxy cluster Abell 1795. Utilizing nearly 750 ks of net ACIS imaging, we are able to resolve the thermodynamic structure of the Intracluster Medium (ICM) on length scales of $\sim 1 \kpc$ near the cool core. We find several previously unresolved structures, including a high pressure feature to the north of the BCG  that appears to arise from the bulk motion of Abell 1795's cool core. To the south of the cool core, we find low temperature ($ \sim 3 \keV$), diffuse ICM gas extending for distances of $\sim 50 \kpc$ spatially coincident with previously identified filaments of H$\alpha$ emission. Gas at similar temperatures is also detected in adjacent regions without any H$\alpha$ emission. The X-ray gas coincident with the H$\alpha$  filament has been measured to be cooling spectroscopically at a rate of $\sim 1 \msolar \yr^{-1}$, consistent with measurements of the star formation rate in this region as inferred from UV observations, suggesting that the star formation in this filament as inferred by its H$\alpha$ and UV emission can trace its origin to the rapid cooling of dense, X-ray emitting gas. The H$\alpha$ filament is not a unique site of cooler ICM, however, as ICM at similar temperatures and even higher metallicities not cospatial with H$\alpha$ emission is observed just to the west of the H$\alpha$ filament, suggesting that it may have been uplifted by Abell 1795's central active galaxy. Further simulations of cool core sloshing and AGN feedback operating in concert with one another will be necessary to understand how such a dynamic cool core region may have originated and why the H$\alpha$ emission is so localized with respect to the cool X-ray gas despite the evidence for a catastrophic cooling flow.

\end{abstract}

\maketitle

\section{Introduction}

Deep, high resolution X-ray observations of the intracluster medium (ICM) in massive galaxy clusters have demonstrated that even apparently ``relaxed'' galaxy clusters are commonly subject to a litany of disruptions near the central cores, which in turn play an important role in understanding the present day thermodynamic structure of the ICM. Even cool core clusters which have remained largely undisturbed on long time scales ($\gtrsim 1 \rm{Gyr}$) have been observed to host a wealth of substructures in their centralmost regions. Cool cores are commonly observed to be subject to disruptions from minor mergers. The signatures of past mergers can be observed long after the ICM phases from the two sub-clusters are apparent and distinct from one another. Many galaxy clusters host spiral shaped substructures, which have been predicted to arise from the ``sloshing'' motions of a cool core \citep[e.g.][]{Ascasibar2006,ZuHone2009} that occur after a cluster merger with a non-zero impact parameter. This offers a second way to uplift low entropy gas from the cool cores of galaxy clusters. Larger scale sloshing motion of cool cores is often observed contemporaneously with the central cluster galaxy hosting an active nucleus (AGN) \citep[e.g.][]{Ehlert2011,Blanton2011}, and the cumulative impact of these two processes operating in tandem may provide even more profound disruptions to the cool core. These sloshing motions have also been suggested to play an important role in suppressing central cooling flows in galaxy clusters without obvious signatures of AGN feedback \citep[see][for reviews of AGN feedback]{McNamara2007,McNamara2012} and may generate significant turbulence in the ICM \citep{ZuHone2013}. One of the ideal targets to study the effects of cool core sloshing along with AGN feedback is the galaxy cluster Abell 1795 ($z=0.062$). This nearby, bright galaxy cluster has a deep ensemble of multiwavelength data, and these observations have shown a wealth of activity near the cool core. In particular, a pair of H$\alpha$ filaments has been observed to extend for distances of $\sim 50 \kpc$ to the south of the BCG \citep[e.g.][]{Cowie1983,Fabian2001,Maloney2001,Crawford2005b,Jaffe2005}. \cite{McNamara1996,Salome2004}, and \cite{McDonald2009} resolved the UV, molecular gas, and H$\alpha$ emission into filaments  rather than a single ``stream''. These H$\alpha$ filaments are also the sites of UV and cold gas \citep[e.g.][]{McNamara1996,Salome2004,McDonald2009,McDonald2012a,McDonald2012b} and bright X-ray emission \citep[e.g.][]{Fabian2001,Ettori2002}. The BCG of Abell 1795 is a powerful radio source with resolved jets observed to its northeast and southwest \citep[4C 26.42, e.g.][]{VanBreugel1984,Ge1993}. Although this cluster has a nominal cooling rate of $\sim 100 \msolar \yr^{-1}$ \citep{Ettori2002}, measurements of the cooling rate in the X-rays and star formation rate in the optical/UV suggest that only $\sim 1-5\%$ of this nominally cooling gas is forming stars \citep{McDonald2009,McDonald2012a,McDonald2012b}. 

In addition to these deep optical and UV studies, extremely deep observations of Abell 1795 ($z=0.062$) with the \cha \ X-ray telescope have been obtained as part of its ongoing calibration efforts, resulting in $\sim 730 \ \rm{ks}$ of net exposure. The only galaxy clusters with similar levels of exposure time are Perseus \citep[][]{Sanders2007,Fabian2008,Fabian2011}, M87 in the Virgo Cluster \citep[][]{Forman2005,Million2010,Werner2010,Werner2012}, and Abell 2052 \citep{Blanton2011}. This extremely deep data set enables a high resolution investigation into the thermodynamic structure of this cluster's ICM on spatial scales approaching the native \cha \ pixel resolution ($0.492 \arcsec$ per pixel, equivalent to $580 \pc$ at Abell 1795's redshift of $z=0.062$) and correlate X-ray structures with small scale features observed at other wavelengths with an unprecedented level of detail. Additionally, we are able to place new and important constraints on the large scale thermodynamic structure of the cluster and its influence on the cool core. 

In this paper we present the first high resolution thermodynamic mapping of Abell 1795 utilizing all of the calibration data available as of August 2013. The structure of this paper is as follows: in Section 2 we discuss our processing of the data, and in Section 3 we discuss features observed in the images. In Section 4 we discuss the methods we use to produce our spectral maps, and in Section 5 we present these maps. In Section 6 we discuss the implications of these results. While this paper focuses on the core regions of Abell 1795, a companion paper discussing the large scale thermodynamic structure of the cluster is currently in preparation. For this study, all cosmological calculations assume a model with $\mysub{H}{0}=71 \km \s^{-1} \Mpc^{-1}$, $\Omegam=0.27$, and $\Omegal=0.73$.

\section{Data Preparation}

A total of 52 \cha \ observations of Abell 1795 were performed using the Advanced CCD Imaging Spectrometer (ACIS) between March 2000 and June 2013. With the exceptions of observations 493 (PI A. Fabian), 494 (PI A. Fabian), and 10432 (PI B. Maughn), all of these observations were taken by the \cha \ calibration team in order to monitor the build-up of molecular contamination on the ACIS filters. The standard level-1 event lists produced by the Chandra pipeline processing were reprocessed
using the CIAO (version 4.5) software package, including the appropriate gain maps and calibration products
(CALDB version 4.5). Bad pixels were removed and standard grade
selections were applied to the event lists. We used the additional
information available in VFAINT mode  to improve the rejection of cosmic ray events whenever the data were taken in that mode. The data were cleaned to
remove periods of anomalously high background using the standard energy ranges and binning methods recommended
by the \cha \ X-ray Center (CXC). Information regarding the observations utilized including their net exposure times after processing are summarized in Table \ref{Observations}.

\begin{table*}
\caption{\label{Observations} A list of the \cha \ observations of Abell 1795 utilized in this study. The columns are: 1) the \cha \ observation ID \# ; 2) the date of the observation; 3) the right ascension of the observation aimpoint, in degrees (J2000); 4) the declination of the observation aimpoint, in degrees (J2000); 5) the net exposure time of the observation after all processing; (6) the distance from the observation aimpoint to Abell 1795's BCG, in arcminutes; and (7) the primary ACIS detector of the observation.        }
\centering

\begin{tabular}{ c c c c  c c c}\\
  \hline 

(1) & (2) & (3) & (4) & (5) & (6) & (7)\\
 Obs ID \# & Date  & RA (aim) & DEC (aim) & Exposure (ks)  & Distance (\arcmin) & Detector \\ 
\hline\hline

493 & 2000-03-21 & 207.2051 & 26.6076 & 19.38 & 1.06$^{\dagger}$ & ACIS-S \\
494 & 1999-12-20 & 207.2349 & 26.6071 & 17.47 & 1.20$^{\dagger}$ & ACIS-S \\
3666 & 2002-06-10 & 207.2037 & 26.5756 & 12.65 & 1.35$^{\dagger}$ & ACIS-S \\
5286 & 2004-01-14 & 207.2293 & 26.6097 & 14.29 & 1.11$^{\dagger}$ & ACIS-S \\
5287 & 2004-01-14 & 207.2292 & 26.6097 & 14.30 & 1.11$^{\dagger}$ & ACIS-S \\
5288 & 2004-01-16 & 207.2288 & 26.6099 & 14.49 & 1.11$^{\dagger}$ & ACIS-S \\
5289 & 2004-01-18 & 207.2294 & 26.6125 & 14.95 & 1.26$^{\dagger}$ & ACIS-I \\
5290 & 2004-01-23 & 207.2534 & 26.7021 & 14.95 & 6.75 & ACIS-I \\
6159 & 2005-03-20 & 207.1369 & 26.6793 & 14.85 & 6.71 & ACIS-I \\
6160 & 2005-03-20 & 207.2058 & 26.6076 & 14.84 & 1.04$^{\dagger}$ & ACIS-S \\
6161 & 2005-03-28 & 207.3310 & 26.5182 & 13.59 & 7.59 & ACIS-I \\
6162 & 2005-03-28 & 207.1998 & 26.6061 & 13.60 & 1.21$^{\dagger}$ & ACIS-I \\
6163 & 2005-03-31 & 207.1984 & 26.6046 & 14.85 & 1.22$^{\dagger}$ & ACIS-I \\
10432 & 2009-03-16 & 207.3580 & 26.4632 & 5.10 & 10.87 & ACIS-I \\
10898 & 2009-04-20 & 207.1952 & 26.5362 & 15.73 & 3.68 & ACIS-I \\
10899 & 2009-04-22 & 207.0909 & 26.5818 & 14.92 & 6.86 & ACIS-I \\
10900 & 2009-04-20 & 207.1963 & 26.5921 & 15.82 & 1.17$^{\dagger}$ & ACIS-S \\
10901 & 2009-04-20 & 207.2007 & 26.4286 & 15.47 & 9.97 & ACIS-S \\
12026 & 2010-05-11 & 207.1953 & 26.5820 & 14.92 & 1.42$^{\dagger}$ & ACIS-I \\
12027 & 2010-03-16 & 207.1473 & 26.6864 & 14.85 & 6.71 & ACIS-I \\
12028 & 2010-05-10 & 207.1979 & 26.5832 & 14.97 & 1.26$^{\dagger}$ & ACIS-S \\
12029 & 2010-04-28 & 207.1965 & 26.5880 & 14.33 & 1.21$^{\dagger}$ & ACIS-S \\
13106 & 2011-04-01 & 207.2004 & 26.6026 & 9.91 & 1.07$^{\dagger}$ & ACIS-S \\
13107 & 2011-04-01 & 207.2003 & 26.6026 & 9.64 & 1.08$^{\dagger}$ & ACIS-S \\
13108 & 2011-03-10 & 207.2097 & 26.6126 & 14.86 & 1.19$^{\dagger}$ & ACIS-I \\
13109 & 2011-03-11 & 207.1635 & 26.6947 & 14.58 & 6.71 & ACIS-I \\
13110 & 2011-03-11 & 207.2095 & 26.6125 & 14.58 & 1.19$^{\dagger}$ & ACIS-I \\
13111 & 2011-03-11 & 207.1614 & 26.6938 & 14.58 & 6.71 & ACIS-I \\
13112 & 2011-03-11 & 207.1938 & 26.6397 & 14.58 & 3.03 & ACIS-I \\
13113 & 2011-03-11 & 207.1778 & 26.6669 & 14.58 & 4.87 & ACIS-I \\
13412 & 2011-05-22 & 207.3265 & 26.6639 & 14.86 & 7.17 & ACIS-I \\
13413 & 2011-05-29 & 207.3188 & 26.6722 & 14.85 & 7.15 & ACIS-I \\
13414 & 2011-05-29 & 207.2490 & 26.5994 & 14.57 & 1.69$^{\dagger}$ & ACIS-I \\
13415 & 2011-05-29 & 207.3283 & 26.6627 & 14.58 & 7.20 & ACIS-I \\
13416 & 2011-05-30 & 207.1734 & 26.5560 & 14.58 & 3.31 & ACIS-I \\
13417 & 2011-06-02 & 207.1353 & 26.5015 & 14.86 & 7.11 & ACIS-I \\
14268 & 2012-03-26 & 207.2049 & 26.6037 & 9.93 & 0.91$^{\dagger}$ & ACIS-S \\
14269 & 2012-04-08 & 207.2006 & 26.5978 & 9.94 & 0.96$^{\dagger}$ & ACIS-S \\
14270 & 2012-03-25 & 207.2010 & 26.6074 & 14.28 & 1.21$^{\dagger}$ & ACIS-I \\
14271 & 2012-03-25 & 207.1484 & 26.6469 & 13.98 & 4.89 & ACIS-I \\
14272 & 2012-03-25 & 207.1218 & 26.6662 & 14.58 & 6.73 & ACIS-I \\
14273 & 2012-03-26 & 207.3146 & 26.5044 & 14.58 & 7.47 & ACIS-I \\
14274 & 2012-04-02 & 207.0984 & 26.6341 & 14.88 & 6.85 & ACIS-I \\
14275 & 2012-04-07 & 207.3479 & 26.5538 & 14.88 & 7.38 & ACIS-I \\
15485 & 2013-04-21 & 207.1963 & 26.5912 & 9.94 & 1.18$^{\dagger}$ & ACIS-S \\
15486 & 2013-04-22 & 207.1962 & 26.5911 & 9.65 & 1.18$^{\dagger}$ & ACIS-S \\
15487 & 2013-06-02 & 207.1997 & 26.5755 & 14.63 & 1.49$^{\dagger}$ & ACIS-I \\
15488 & 2013-04-10 & 207.1284 & 26.6140 & 14.58 & 4.95 & ACIS-I \\
15489 & 2013-04-08 & 207.0970 & 26.6268 & 14.58 & 6.78 & ACIS-I \\
15490 & 2013-04-17 & 207.3536 & 26.5730 & 14.89 & 7.39 & ACIS-I \\
15491 & 2013-04-18 & 207.0896 & 26.5813 & 14.89 & 6.93 & ACIS-I \\
15492 & 2013-04-15 & 207.3540 & 26.5801 & 14.58 & 7.34 & ACIS-I \\

&&&&&\\
\hline\hline

\end{tabular}
\end{table*}

\section{Imaging Analysis}\label{imaging}

\begin{figure*}
\includegraphics[width=0.95\textwidth]{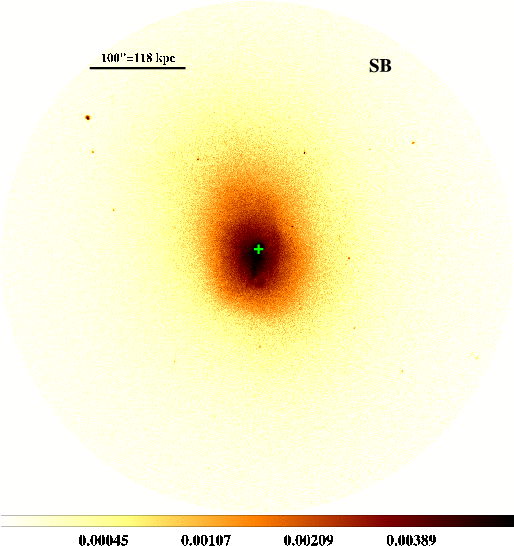}
\caption{\label{WideFieldSB}Background subtracted, exposure corrected \cha \ image of Abell 1795 in the energy band of $0.5-8.0 \keV$ utilizing all 730 ks of net exposure, in units of $\rm{photons} \cm^{-2} \rm{pixel}^{-1}$. The green cross denotes the position of the BCG.   }
\end{figure*}

\begin{figure*}
\centering
\subfigure[]{
\includegraphics[width=0.47\textwidth]{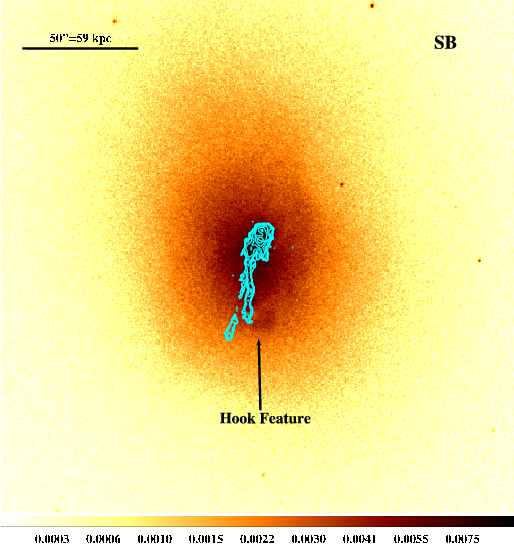}
\label{CoreSB}
}
\subfigure[]{
\includegraphics[width=0.47\textwidth]{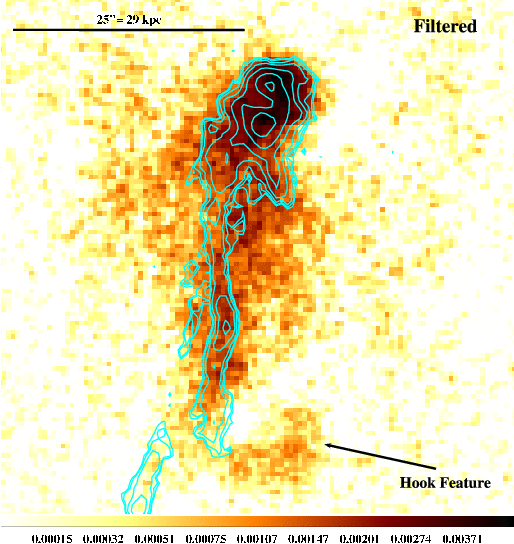}
\label{CoreSBFilt}
}
\subfigure[]{
\includegraphics[width=0.47\textwidth]{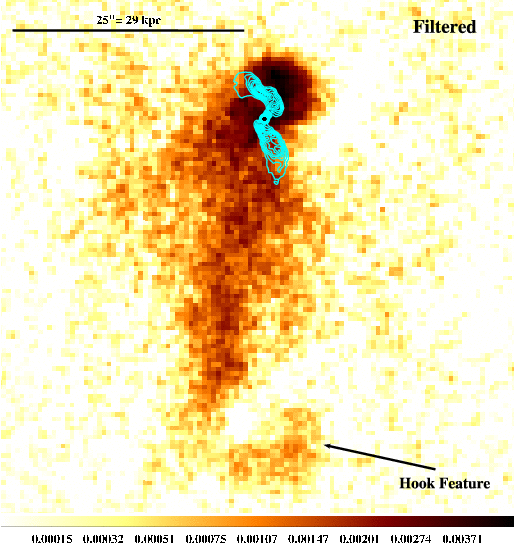}
\label{ZoomCoreSBFilt}
}

\caption{  Small scale features in the core of Abell 1795. {\it Top:} The same image as Figure \ref{WideFieldSB}, zoomed in to focus on smaller scale structures near the cool core. Overlaid in cyan are the contours of the previously identified H$\alpha$ filament \citep[e.g.][]{McDonald2009}. The ``hook'' is observed just to the west of the southern tip of the H$\alpha$ filament and labeled. In order to optimize the spatial resolution, this image only includes observations with aimpoints within $2 \arcmin$ of the BCG, which constitutes $\sim 342 $ ks of exposure time. {\it Bottom Left:} The resulting surface brightness image after applying a high frequency bandpass filter to the image in \ref{CoreSB}. The ``hook'' just to the west of main tail is more apparent here as well as the apparent break in surface brightness between the ``hook'' and the neighboring H$\alpha$ filament. Overlaid in cyan are the contours of H$\alpha$ emission. It is clear that the tail of dense X-ray gas is significantly wider than the H$\alpha$ filament to the west, although the H$\alpha$ is located at the site of the brightest X-ray emission. {\it Bottom Right:} The same image as in \ref{CoreSBFilt}, but with 1.4 GHz radio contours from \cite{Ge1993} overlaid in cyan. Depressions in the X-ray surface brightness are observed coincident with the radio jets, although these two cavities are the only two definitively observed in this image.      
}
\label{PubImage}
\end{figure*}

The final background subtracted, exposure-map corrected image of Abell 1795 is shown in Figure \ref{WideFieldSB}. To produce this image, we first determined the background for each observation  utilizing blank sky event files provided by the CXC. We then co-added the 52 net count images and exposure maps to produce co-added counts images and exposure maps. We then divided the final counts image by the final exposure map to produce the final background subtracted, exposure corrected image. Images that focus on the substructures within the cool core are shown in Figure \ref{PubImage}, which shows the surface brightness image in Figure \ref{CoreSB} as well as a high pass filtered image of Abell 1795 to highlight small scale structures in Figures \ref{CoreSBFilt} and \ref{ZoomCoreSBFilt}. To produce the high pass filtered image, we performed a Fourier transform on the image, multiplied the transformed image by the function

\begin{equation}
F(k)=\frac{\left(\frac{k}{\mysub{k}{0}}\right)^{2}}{1+\left(\frac{k}{\mysub{k}{0}}\right)^{2}}
\end{equation}
\noindent where $k$ is the distance from the image origin in frequency space,  and then transformed the image back into real space. For this image, we have set the scale length $\mysub{k}{0}=(2\pi/15$ \arcsec).

These imaging data reveal several clear features. The most notable features are located near the core of the cluster. The X-ray surface brightness morphology traces the shape of the H$\alpha$ filament extremely well. While the H$\alpha$ filaments appear to correlate with the very brightest component of the structures to the south of Abell 1795's BCG, diffuse gas apparently associated with these filaments at similar surface brightness is observed to the west of the H$\alpha$ emission. We also identify a hook shaped feature just to the west of the southern tip of the H$\alpha$ filament, previously identified in \cite{Crawford2005b}. There appears to be a break in the surface brightness between the hook and the main filament, which makes it unclear as to whether or not these two structures are connected physically to one another. Visual inspection of these images does not show any compelling evidence of cavities in the X-ray beyond the two located in the immediate vicinity of the radio jets (see Figure \ref{ZoomCoreSBFilt}), with the possible exception of the region just to the north of the hook, which could plausibly denote the boundary of a cavity. However, as discussed in \cite{Crawford2005b} there are no optical or radio counterparts associated with this region, and its origin remains unclear. Additional larger scale radio observations will be required to determine whether this cavity is coincident with radio emission consistent with jets. It also is clear that there is a distinct edge in surface brightness just to the north of the BCG in the bandpass filtered image.

\section{Spectral Analysis}\label{spectra}
\subsection{Methods}\label{spectralmethods}
The \cha \ observations of Abell 1795 enable us to carry out detailed, spatially 
resolved measurements of the thermodynamic quantities of the ICM at high spatial resolution. All
spectral analysis was carried out using {\small XSPEC} \rm   \citep[][version 12.5]{Arnaud2004}.

\subsubsection{Regions of Interest}\label{spectralbins}

Spatially resolved spectral maps for the cluster were extracted in regions determined by the contour 
binning algorithm of \citet{Sanders2006}, which 
creates bins of equivalent signal-to-noise, following contours in surface brightness. 

We have produced our default thermodynamic maps for this study at a signal-to-noise of 100 (approximately 10,000 counts per bin across all observations). We have also produced maps at a signal-to-noise of 50 to offer even higher spatial resolution maps immediately surrounding the BCG. We fit all of the spectra for each bin simultaneously with our chosen model, as the different aimpoints lead to sizable observation-dependent variations in the \cha \ detector response at a particular sky position.

\subsubsection{Modeling the Emission}

All spectral regions were initially modeled as a single temperature optically thin plasma using the {\small MEKAL} \rm  code 
of \citet{Kaastra1993} incorporating the
Fe-L calculations of \citet{Liedhal1995} and the photoelectric absorption models of \citet{McCammon1992}. 
We used the determinations of 
solar element abundances given by \citet{Lodders2003}.
The abundances of metals ($Z$) were assumed to vary with a common ratio with respect to the Solar
values. The single-temperature plasma model has three free parameters: the temperature ($kT$), the metallicity ($Z$), 
and normalization ($N$).

In each region, the spectral analysis assumed a fixed Galactic absorption column of 
$1.21 \times 10^{20} \cm^{-2}$ \citep{Kalberla2005}. 
The modified Cash statistic in {\small XSPEC} \rm \citep{Cash1979,Arnaud2004} was minimized to determine the best fit 
model parameters and uncertainties. All uncertainties
given are 68\%  ($\Delta C=1$) confidence intervals, unless otherwise noted.  We account for the background in our spectral fits using both blank-sky \cha \ observations provided by the CXC, renormalized such that the count rates between the Abell 1795 observations and the blank sky files are the same in the $9.5-12 \keV$ energy band. 
 
Whenever we perform a deprojected fit, we used the {\small PROJCT} mixing model in {\small XSPEC}, which simultaneously fits models to the spectra from multiple annuli while accounting for the geometric contributions of one annular shell into neighboring shells. Since the uncertainties between the temperature/density measurements in deprojected spectral fits are correlated with one another, we determine statistical uncertainties on model parameters using a Markov-Chain Monte Carlo (MCMC) analysis within the {\small XSPEC} framework.

\subsubsection{Thermodynamic Quantities}
Several thermodynamic quantities can be calculated directly from the
best fit {\small MEKAL} \rm model parameters. The 
electron density ($n_{e}$), pressure ($P$), and entropy ($K$) of 
the ICM are derived from the {\small MEKAL} \rm temperature ($kT$) and normalization
($N$) as
\begin{equation}\label{NormDensity}
\mysub{n}{e}^{2}=\frac{4\pi \times 10^{14}\left(1+z\right)^{2}\mysub{D}{A}^{2}N}{1.2V}
\end{equation}

\begin{equation}
P=kT\mysub{n}{e}
\end{equation}

\begin{equation}
K=kT\mysub{n}{e}^{-2/3}
\end{equation}

\noindent where the cosmological value of the angular diameter distance $\mysub{D}{A}$ is 243 \Mpc \ 
at the cluster redshift. The volume of the region was estimated as

\begin{equation}
V=\mysub{D}{A}^{3}\Omega \sqrt{\mysub{\theta}{max}^{2}-\mysub{\theta}{min}^{2}}
\end{equation}
where $\mysub{\theta}{max,min}$ are the maximum and minimum angular distances of any point in the region 
 to the center of the cluster, respectively, and $\Omega$ is the solid-angle 
extent of the region in the sky \citep{Henry2004,Mahdavi2005}. For deprojected regions, we calculate the volume as the volume of the spherical shell from which the spectrum was extracted.

Since the uncertainties on the temperature are considerably larger than those on the density ($\sim 5-10\% $ for the 
temperature as compared to $\sim 1\%$  for the densities), the fractional uncertainties on the pressure 
and entropy are similar to the corresponding 
temperature measurement.

\section{Spectral Results}\label{thermodyn}

\subsection{Temperature Structure }

\begin{figure*}
\centering
\subfigure[]{
\includegraphics[width=0.47\textwidth]{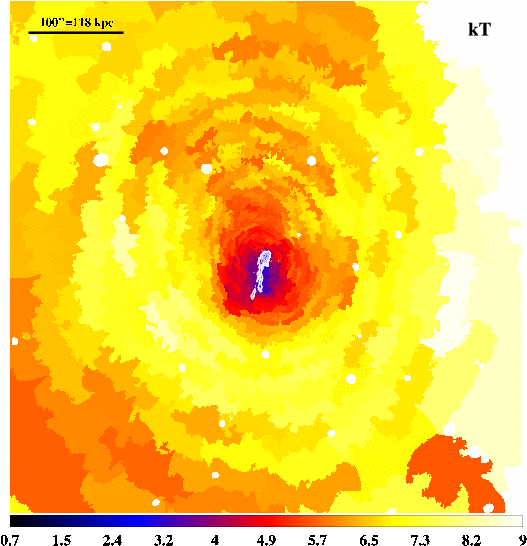}
\label{TempWide100}
}
\subfigure[]{
\includegraphics[width=0.47\textwidth]{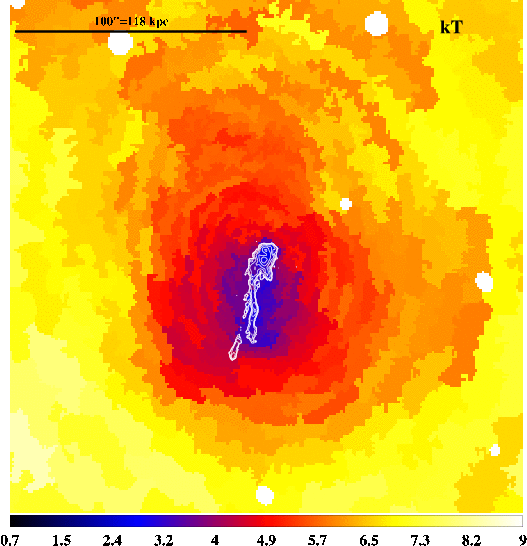}
\label{TempCore100}
}
\subfigure[]{
\includegraphics[width=0.47\textwidth]{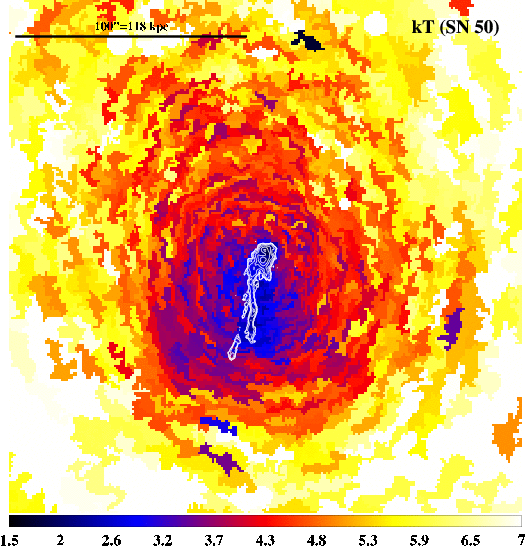}
\label{TempCore50}
}
\caption{ Projected temperature structure of the ICM in Abell 1795, in units of \keV. In all three images the white contours denote the previously observed H$\alpha$ filaments. {\it Top:} Wide field temperature map of Abell 1795, using a constant signal-to-noise of 100. The measurement uncertainties are at the $\sim 5-10\%$ level at this signal-to-noise.  {\it Lower Left:} The same temperature map as the top image, zoomed in to show the region immediately surrounding the BCG. Cool gas is spatially coincident with the H$\alpha$ filaments, but cool gas extends further in the east-west direction than H$\alpha$ emission.  {\it Bottom Right:} The same temperature map as the lower left image binned with a lower signal-to-noise threshold of 50. It is especially clear in this image that the cool tail is both spatially resolved and significantly wider than the H$\alpha$ filaments.    }
\label{TempMaps}
\end{figure*}

\begin{figure}
\includegraphics[width=\columnwidth]{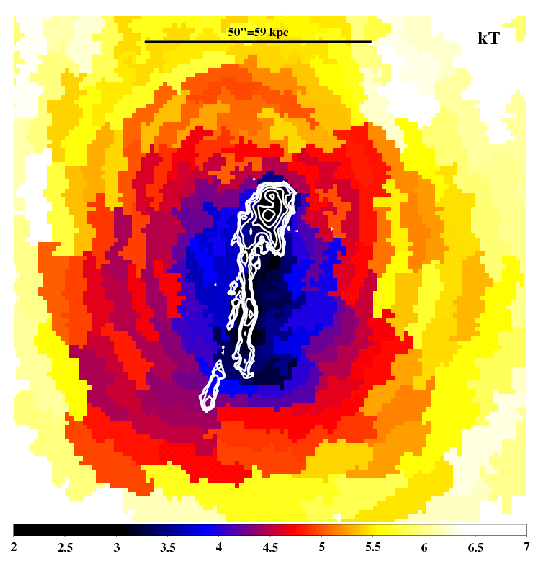}
\caption{\label{CoreTemp}  The same temperature map of Figure \ref{TempCore100} further zoomed in to show the temperature structure immediately surrounding the cool core. It is clear from this figure that the tail of $\sim 2-3 \keV$ ICM is significantly wider than the H$\alpha$ filament and that there is a sharp edge in the temperature structure immediately to the north of the BCG.        }
\end{figure}

Temperature maps for Abell 1795 are shown in Figure \ref{TempMaps}. These maps show a clear tail of cool ($\sim 2-3 \keV$) gas trailing to the south of the brightest gas and BCG. It is clear that the H$\alpha$ filaments are spatially coincident with this cool gas. However, it is clear from both temperature maps that the cool gas is not uniquely situated to the H$\alpha$ filaments, as spatially resolved regions at similar temperatures of $\sim 2-3 \keV$ are detected immediately to the west of the H$\alpha$ filament. This is especially clear in our SN50 map (Figure \ref{TempCore50}), where there are more than 10 independent regions to the west of the H$\alpha$ filaments that are measured to have similar temperatures as the gas within the H$\alpha$ filaments. To more clearly differentiate the temperature structure of the cool core, we present a higher zoom and contrast temperature map in Figure \ref{CoreTemp} designed to present temperature variations in the vicinity of the H$\alpha$ filaments. 

\begin{figure}
\includegraphics[width=\columnwidth]{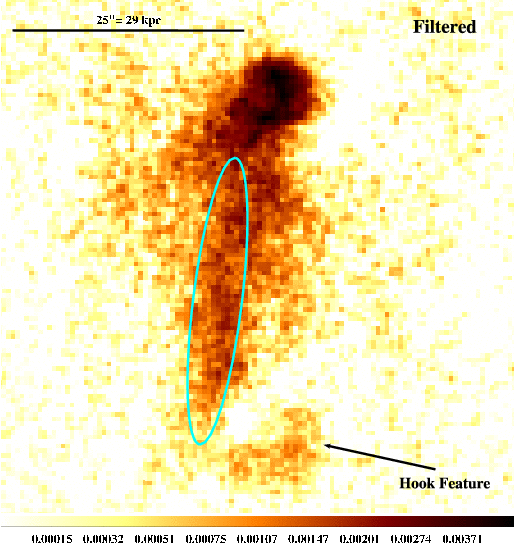}
\caption{\label{TailReg}  Bandpass filtered image of Abell 1795 that shows the elliptical region we used to measure the spectroscopic cooling rate within the H$\alpha$ filament. Within the locus of the cyan ellipse, we find significant evidence for spectroscopic cooling of the X-ray gas at a rate of $\sim 1 \msolar \yr^{-1}$.      }
\end{figure}

In order to understand the extent to which we can measure a cooling flow at the site of the H$\alpha$ filaments, we extract the X-ray spectrum from this region and fit it with a cooling flow model ({\small PHABS(MEKAL+MKCFLOW)} in {\small XSPEC}). The specific region we chose is shown in Figure \ref{TailReg}. The lower temperature of the cooling flow model was fixed to $0.1 \keV$, while the upper temperature was tied to the {\small MEKAL}'s temperature. The metallicities of the two components were also tied together. Across the entire region where the H$\alpha$ filament is detected we measure a cooling flow of $\dot{M}= 3.32 \pm 0.24 \msolar \yr^{-1}$. Excluding the core region immediately surrounding the BCG we measure a cooling flow rate of  $\dot{M}= 1.08 \pm 0.19 \msolar \yr^{-1}$. This latter result is in good agreement with estimated star formation rates derived from UV observations of this region, which imply an extinction corrected star formation rate of $\mysub{\dot{M}}{SFR} =1.4 \pm 0.2 \msolar \yr^{-1}$ in the filament (excluding the region surrounding the BCG) assuming a Salpeter IMF \citep{McDonald2011a}.\footnote{Assuming an IMF that is not as ``top-heavy'' as a Salpeter results in a star formation rate of $\mysub{\dot{M}}{SFR} =0.7 \pm 0.2 \msolar \yr^{-1}$, still in good agreement with the X-ray cooling flow.} Based on the changes in the $C$-statistic as compared to a single-temperature model  ($\Delta C =31$ for 1 additional degree of freedom), the addition of a cooling flow is a statistically significant improvement to the fit. The overall value of the $C$-statistic for this fit is $22961$ with $25596$ degrees of freedom, suggesting that is a formally a ``good'' fit to the data. 

To the north of the BCG, there is little evidence for gas at temperatures below $\sim 4 \keV$ beyond the sharp surface brightness edge, suggesting this edge is in fact a cold front resulting from the cool core's bulk motion northward.

\begin{figure*}
\centering
\subfigure[]{
\includegraphics[width=0.47\textwidth]{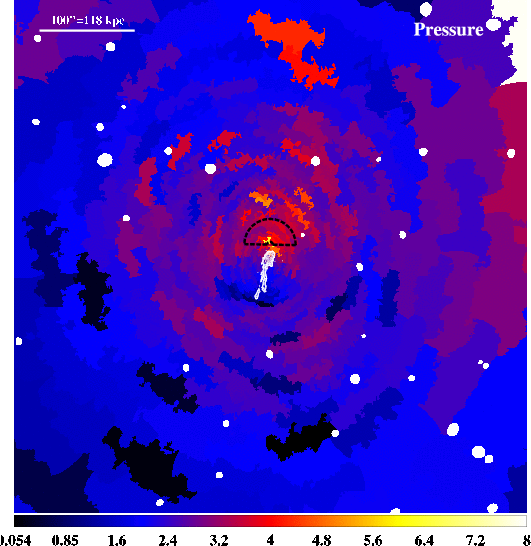}
\label{PressureWide}
}
\subfigure[]{
\includegraphics[width=0.47\textwidth]{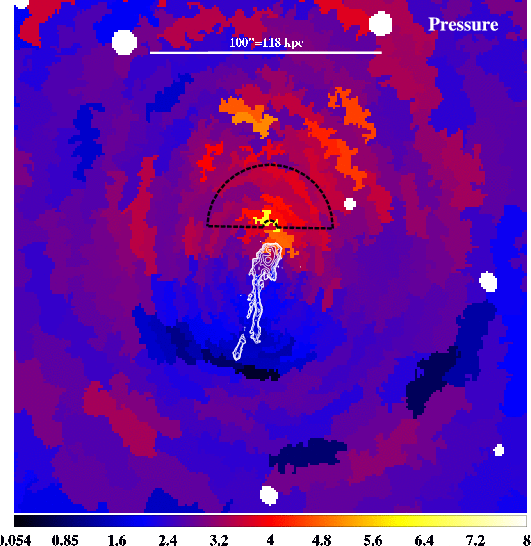}
\label{PressureCore}
}
\subfigure[]{
\includegraphics[width=0.47\textwidth]{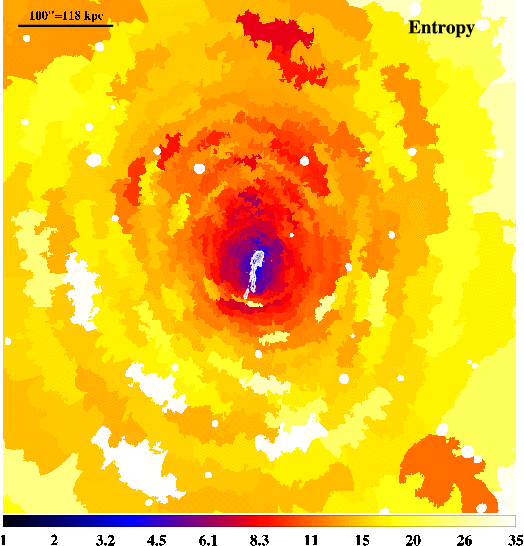}
\label{EntropyWide}
}
\subfigure[]{
\includegraphics[width=0.47\textwidth]{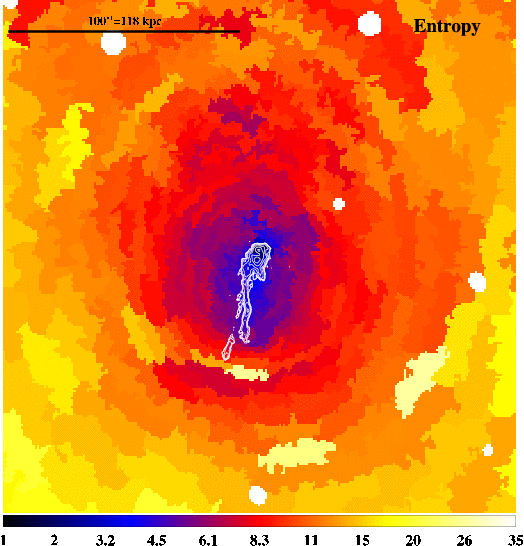}

\label{EntropyCore}
}
\caption{\label{PKMaps}Projected pressure (top row) and entropy (bottom row) structure of the ICM in Abell 1795, in units of $\keV \cm^{-3}$ and $\keV \cm^{2}$, respectively. As in the previous figures, the white contours denote the position of the H$\alpha$ filaments. Uncertainties in the pressure and entropy are approximately $\sim 5-10 \%$. {\it Top Left: }The wide field pressure map of Abell 1795. {\it Top Right: } The pressure structure near the central core of Abell 1795. Anomalously high pressure region is detected approximately $5.5 \arcsec (\sim 6 \kpc)$ to the north of the cool core, the approximate location of which is denoted by the black, dashed, annular wedge. It is also clear from this image that the H$\alpha$ filament has the same apparent thermal pressure as its surroundings. {\it Bottom Left: } The wide field entropy structure of Abell 1795.  {\it Bottom Right: } The entropy structure in the core of Abell 1795. While the H$\alpha$ filament is coincident with the lowest entropy gas, the ICM immediately surrounding the H$\alpha$ tail has similar entropy values. There is also a clear north-south asymmetry to the entropy structure, with the entropy gradients being much sharper to the south than to the north.      }
\end{figure*}

\subsection{Pressure Structure} 

The projected pressure structure of Abell 1795 is shown in Figures \ref{PressureWide} and \ref{PressureCore}. Both of these maps show that the filament observed in the temperature structure is not readily apparent in its pressure, suggesting that this filament region is in approximate thermal pressure equilibrium with its surroundings, at least in projection. 

\begin{figure*}
\centering
\subfigure[]{
\includegraphics[width=0.47\textwidth]{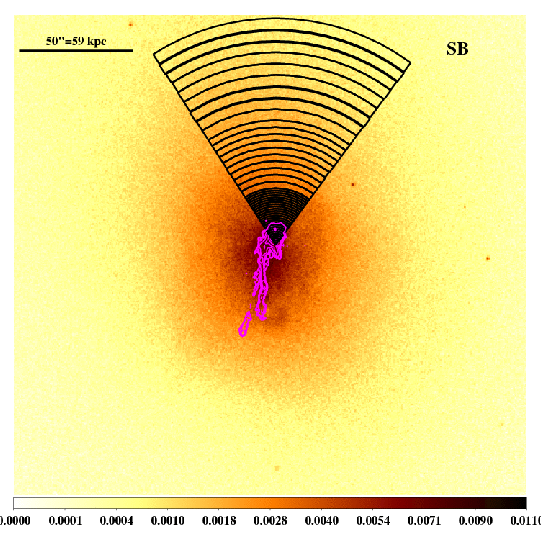}
\label{SBRegionsNorth}
}
\subfigure[]{
\includegraphics[width=0.47\textwidth]{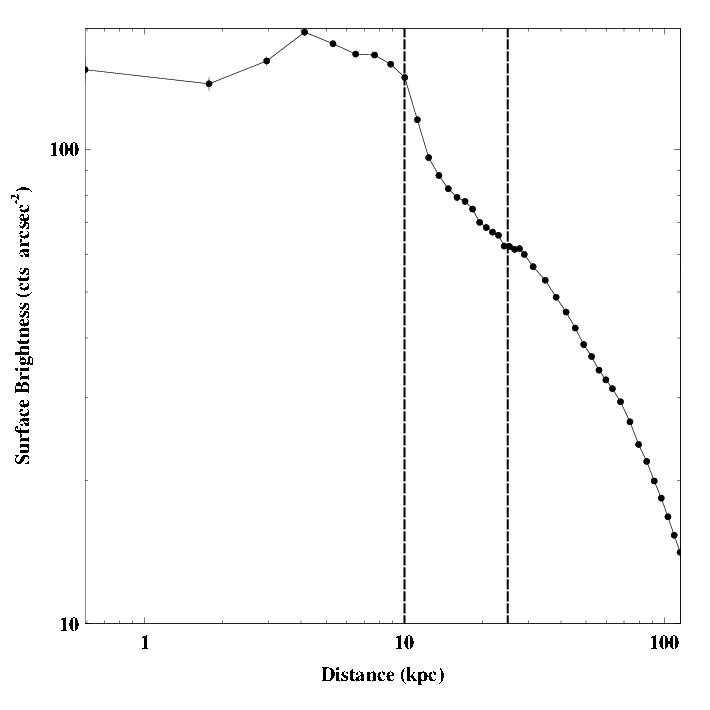}
\label{SBProfNorth}
}
\caption{\label{SBNorth} Surface brightness profile of Abell 1795 to the north of the BCG, in the direction of the candidate shock front. {\it Left:} The X-ray image of Abell 1795's core overlaid (in black) with the regions used to determine the surface brightness profile across the shock front. {\it Right:} The surface brightness profile (in units of $\rm{cts} \ \rm{arcsec}^{-2}$ as a function of distance from the BCG in \kpc. The inner and outer vertical dashed lines denote the boundaries of the cold front and the candidate shock front, respectively. It is clear that at distances of $\sim 30 kpc$ from the BCG that there is a sharp break in the slope of the surface brightness profile suggestive of compression by a shock front.            }
\end{figure*}

\begin{figure*}
\centering
\subfigure[]{
\includegraphics[width=0.47\textwidth]{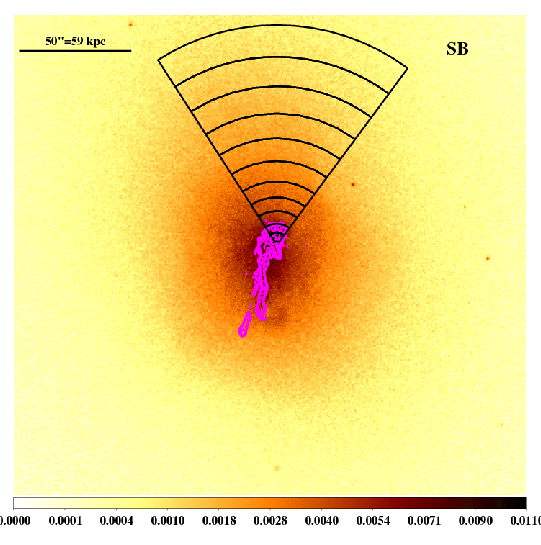}
\label{RegionsNorth}
}
\subfigure[]{
\includegraphics[width=0.47\textwidth]{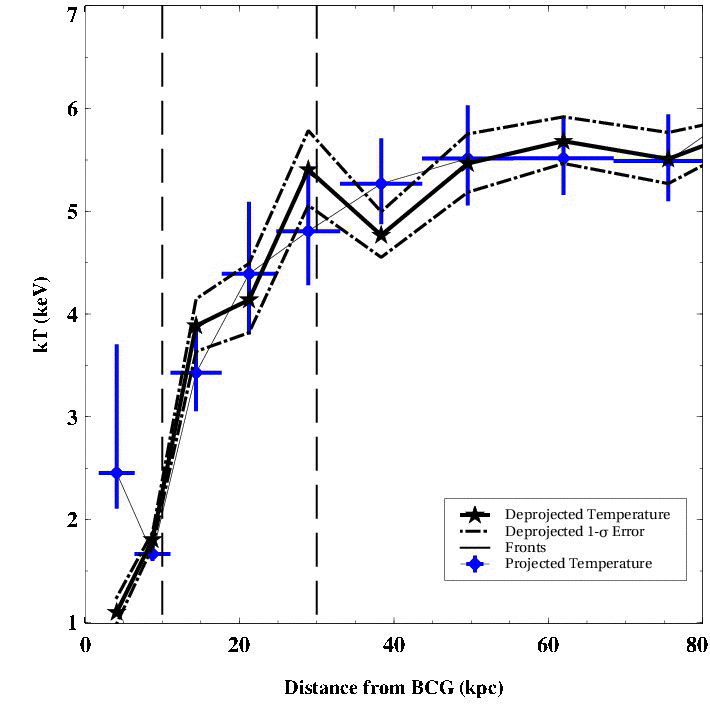}
\label{TempProfNorth}
}
\subfigure[]{
\includegraphics[width=0.47\textwidth]{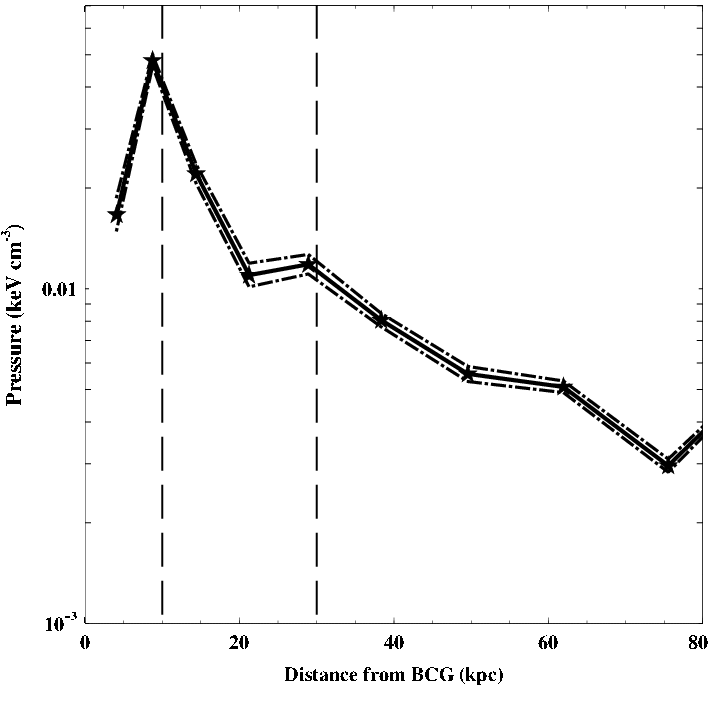}
\label{PressProfNorth}
}
\caption{\label{PressureNorth} \scriptsize Temperature and pressure structure to the north of the BCG, the same direction as the cold front. The inner and outer dashed vertical lines denote the positions of the cold front edge and candidate shock front detected in the surface brightness profile, respectively. The dashed curves denote the 68\% confidence interval for the fit parameters, and the stars denote the centers of the annular regions utilized to determine these profiles.  {\it Top}: The X-ray image of the core of Abell 1795 overlaid (in black) with the regions used to determine the deprojected temperature profile across the candidate shock front. {\it Bottom Left:} Projected (blue crosses) and deprojected temperature profile (black solid and dashed lines) along the northern direction, as a function of distance from the BCG. The clear jump in temperature beyond the cold front is apparent. A small jump in the deprojected temperature is observed in the candidate shock front region that is not observed in the projected temperature profile. {\it Bottom Right:} The deprojected pressure profile along the north of Abell 1795. The high pressure jump associated with the cold front is clear, and is roughly a factor of two larger than the observed pressure outside of the cold front region. A break in the pressure profile is also observed at the site of the candidate shock front, but deeper data will be necessary to confirm the presence of a shock at this location.           }
\end{figure*}

There is a clear detection of a north-south asymmetry in the pressure structure of the ICM surrounding the cool core. As shown in Figure \ref{PressureCore}, significantly higher pressures are measured immediately to the north of the cool core compared to the azimuthal average. The highest pressure is located well to the north of the X-ray surface brightness peak (which is also the site of the BCG and the brightest H$\alpha$ emission). We further examine the thermodynamic structure of the ICM near this feature by extracting the profiles for the surface brightness, projected and deprojected temperature, and pressure across the northward direction, the results of which are shown in Figures \ref{SBNorth} and \ref{PressureNorth}. The surface brightness profile shows two very clear breaks in its slope: the first of these breaks is associated with a region of low temperature ($\sim 1.5 \keV$) ICM which abruptly transitions to much higher temperatures ($\sim 3.5 \keV$). Such a feature suggests the presence of a cold front \cite{Markevitch2007} just to the north of the BCG, the origin of which is the northward motion of the cool core. The second of these surface brightness breaks occurs approximately 20 \kpc \ further to the north of the cold front. The sudden flattening of the surface brightness in this region is suggestive of a shock front. 
We can determine an estimate for the velocity of the cold front using the measured pressure difference across the cold front as an estimate of the ratio for the free stream pressure ($\mysub{p}{1}=0.022 \keV \cm^{-3}$, radial bin \#3 in Figure \ref{PressProfNorth}) and at the stagnation point ($\mysub{p}{0}=0.048 \keV \cm^{-3}$, radial bin \#2 in Figure \ref{PressProfNorth}), which allows us to calculate the Mach number associated with its motion as \citep{Landau1959,Markevitch2007}

\begin{equation}
\frac{\mysub{p}{0}}{\mysub{p}{1}}=\left(\frac{\gamma+1}{2}\right)^{(\gamma+1)/(\gamma-1)} \mach^{2}\left[ \gamma- \frac{\gamma-1}{2\mach^{2}}\right]^{-1/(\gamma-1)}
\end{equation}
where $\mach$ \ is the Mach number in the free stream and $\gamma=5/3$ is the adiabatic index of the ICM gas. We calculate that for the observed pressure jump of $\mysub{p}{0}/\mysub{p}{1} \sim 2.2$ that we expect a Mach number of $\mach \sim 1.35$. Because a fraction of this pressure difference arises from the differences in the gravitational potential at these different radii, this represents an upper limit as to the Mach number of the cold front's motion\footnote{This particular formula for the pressure jump across the cold front implicitly assumes $\mach > 1$, and a different formula is used for $\mach < 1$. However, when we calculate the expected sub-sonic Mach number it also implies supersonic motion, suggesting that the pressure jump across the cold front is simply too large to arise from a subsonic cold front.}. We cannot, however, confirm the presence of this shock using the measured temperature and pressure profiles. While the deprojected temperature and pressure profiles do indeed show some weak evidence for shock heating at the site of the second surface brightness jump, this jump does not appear in the projected temperature profile, suggesting that it is most likely an artifact of our deprojection procedure. Deeper data that is able to resolve the temperature structure on smaller length scales better suited to its width (as measured in the surface brightness profile) will be required to confirm the presence of a shock at this location.

\subsection{Entropy Structure} 

Projected entropy maps for Abell 1795 are found in Figures \ref{EntropyWide} and \ref{EntropyCore}. While the H$\alpha$ filament region is coincident with the lowest entropy gas, it is not the exclusive site of such low entropy gas. Similar to what is observed in the temperature maps, ICM gas at similarly low entropies is also observed to the west of the H$\alpha$ filament. The entropy of the tail region is considerably lower than the ambient ICM at similar clustercentric distances to the north, however, consistent with \cite{McDonald2010}. 

The entropy structure of the cool core region has a high degree of asymmetry in the north-south direction. In particular, the entropy gradient in the southward direction is much steeper and more pronounced than to the north, suggesting that there may have been some additional source of heating to the north of the BCG. Such an asymmetry is consistent with what is observed in the surface brightness profile of the cluster at larger scales with \suz \ \citep{Bautz2009}, where the cluster is noticeably brighter to the north than to the south.  

\subsection{Two-Temperature Fits and Metallicity Structure}

\begin{figure*}
\centering
\subfigure[]{
\includegraphics[width=0.47\textwidth]{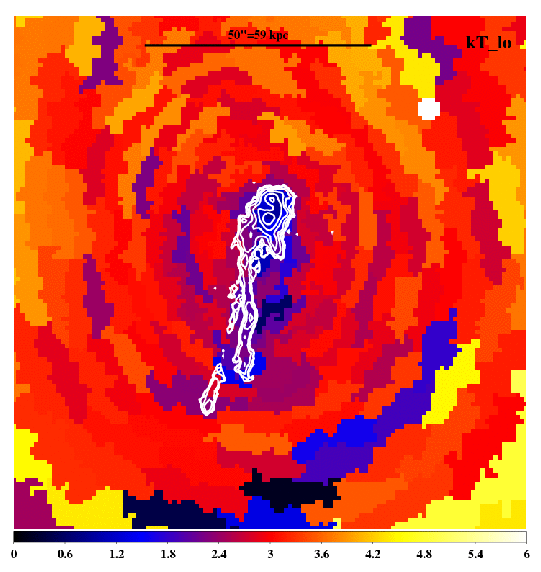}
\label{kTlo}
}
\subfigure[]{
\includegraphics[width=0.47\textwidth]{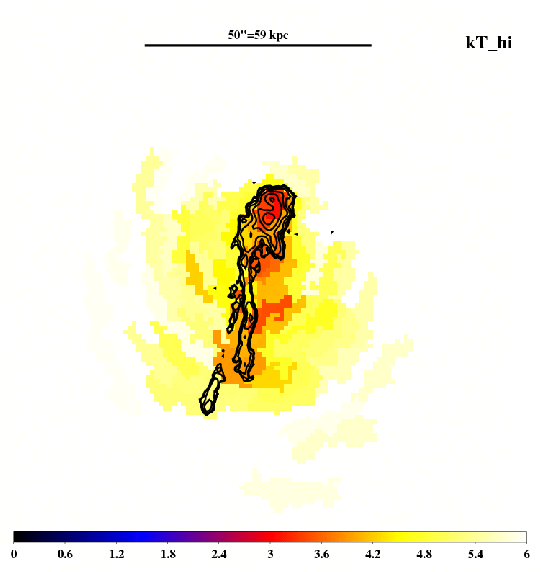}
\label{kThi}
}
\caption{\label{TwoTemp} Temperature structure in the core of Abell 1795 using a two-temperature model fit. These two figures have identical color bars. {\it Left:} The lower of the two temperatures in our two-temperature fit. {\it Right:} The higher of our two temperatures. Regions in white have no significant emission measure at temperatures below $\sim 6 \keV$ and do not require more than a single temperature to account for their spectra. These maps show that the ICM just to the south of the BCG reaches temperatures as low as $\sim 1.8 \keV$ and is equally distributed between the H$\alpha$ filament and the adjacent region without H$\alpha$ emission.                 }
\end{figure*}

\begin{figure}
\includegraphics[width=\columnwidth]{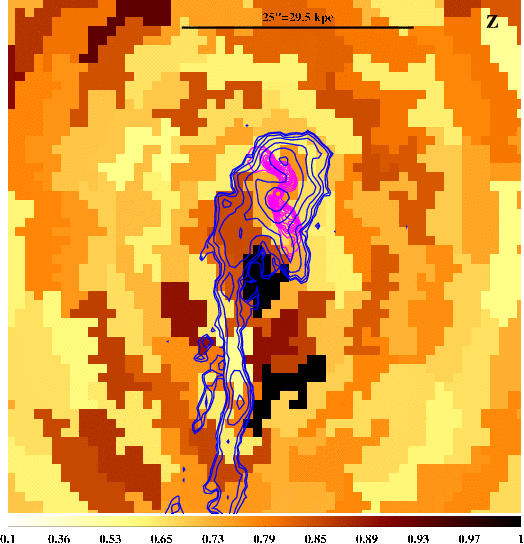}
\caption{\label{MetalsMap} Metallicity structure of the core region of Abell 1795, in solar units. The metallicity was determined for each region by fitting the SN100 data with a two-temperature models whose metallicities were tied together. Overlaid in blue are the H$\alpha$ filament contours while the radio contours are in magenta. Uncertainties in the metallicities with these fits are at the $\sim 15 \%$ level. The highest metallicity gas is located immediately to the south of the radio lobes and to the west of the H$\alpha$ filament, while the gas immediately surrounding the BCG is poorer in metals than the surrounding ICM, suggesting that metal-rich gas may have been uplifted by the central outburst.   }
\end{figure}

We show the metallicity map of the ICM for Abell 1795 in Figure \ref{MetalsMap}. For this map, each SN100 bin was fit with a two-temperature model where both phases have a common metallicity. The two-temperature fits are shown in Figure \ref{TwoTemp}, and are generally consistent with the results of the single-temperature model fits. These results further confirm that gas at similar temperatures are found in both the H$\alpha$ filament and immediately to the west of the filament, and suggest the ICM in the H$\alpha$ filament can reach as low as $\sim 1.8 \keV$.  Uncertainties on the metallicity from these fits are at the $\sim 15-20\%$ level. As this figure shows, the metallicity near the core of Abell 1795 is consistently measured to have a value of $Z \sim 0.6-0.7 \zsolar$. One structure of interest is immediately to the south of the western radio lobe, where a region of higher ($Z \sim  \zsolar$) metallicity is observed, which may be associated with the uplift of metals from the BCG region. This is further suggested by the relatively low metallicity ($Z \sim 0.3 \zsolar$) immediately surrounding the BCG. Observations of AGN feedback in other clusters have shown that metal rich regions are often found coincident with the edges of radio lobes, likely from the uplift of metal-rich gas by the AGN jets \citep[e.g.][]{Simionescu2009,McNamara2012}. We emphasize that this high metallicity is coherently measured across three independent regions in our map, and not simply based on the measurement in a single region.

\section{Discussion}

\subsection{The Origin of the H$\alpha$ Filaments}

{The \cha \ observations clearly show ICM at temperatures of $\sim 3 \keV$ and high densities separated from the BCG. After escaping the sphere of influence of central AGN feedback, this ICM should cool down to star-forming temperatures ($\sim 10-100 \K$) on short time scales. With these data we are able to robustly measure the X-ray cooling flow in the H$\alpha$ filament, which shows that the spectroscopic cooling rate for this region is in good agreement with the UV derived star formation rate. Such a connection suggests that these H$\alpha$ filaments may be primarily powered by star formation that results from the cooling flow of X-ray gas down to star forming temperatures. This scenario is also consistent with the line ratios measured in \cite{McDonald2012b}, as there is sufficient UV emission to account for all of the observed H$\alpha$ emission in this filament \citep{McDonald2009}. A similar spatial coincidence between cool X-ray emitting gas and H$\alpha$ filaments has been seen in a stripped tail behind the galaxy ESO 137-001 falling into the cluster Abell 3627 \citep{Sun2007}. The stripped gas trailing behind ESO 137-001 is a site of star formation taking place outside of the galaxy within the ICM, and our data are consistent with this same picture occurring just to the south of Abell 1795's BCG. It is not clear from these observations alone whether or not such a scenario of stripping+radiative cooling is common in galaxy clusters, and high-resolution studies of other clusters will be necessary to understand how unique this physical situation is. Since the H$\alpha$ tail extends out in the opposite direction of the observed radio jets, these data disfavor a scenario where the central AGN outburst is responsible for these filaments. There is also no compelling evidence that cavities in the X-ray emission from a previous outburst are detected on the length scales of the filament. One potential site for a larger scale cavity is the hook shaped feature located just to the west of the H$\alpha$ filament's southern tip, although larger scale radio observations at lower frequencies will be necessary to associate this feature with a previous AGN outburst, given that these \cha \ observations show no other obvious signatures for this feature being a cavity.

While the data are consistent with a scenario where the coolest, densest X-ray emitting gas in the H$\alpha$ filament region eventually cools to star forming temperatures, it is not clear why only a fraction of the coolest X-ray gas appears to host star formation (i.e. why the H$\alpha$ filament is so localized as compared to the $\sim 2-3 \keV$ tail). All of the gas in this tail is at similar temperatures and entropies, so naively one expects this entire tail to cool homogeneously. Small scale physics of the ICM such as turbulence or magnetic fields operating differently in the eastern and western halves of this tail is the likely the source for such differences.  Turbulence does not appear to be a significant contributor in the eastern half of the X-ray tail given the narrow, linear structure of the H$\alpha$ filament, but the data do not provide any clues regarding the structure of turbulence in the western half. It is possible that the central AGN outburst may be a source of turbulent heating and mixing in the western half of the tail. We find that the cool X-ray gas begins lacking H$\alpha$ emission just to the south of the observed radio jets. This particular ICM gas (the cool ICM not associated with the H$\alpha$ filament to the west of the H$\alpha$ filament) also appears to be especially rich in metals, suggesting that it may have been uplifted from the BCG by AGN feedback similarly to what is observed in clusters such as Hydra A and MS 0735.6+7421\citep[e.g.][]{Simionescu2009,McNamara2012}. The uplift of this gas initially located around the BCG may provide sufficient turbulence to suppress the collapse of X-ray gas that would otherwise cool down to star forming temperatures. This would also account for the needed second source of heating and mixing required by the measured ratios of \ion{O}{1}/H$\alpha$ and \ion{N}{2}/H$\alpha$ observed here \citep{McDonald2012b}, and is also consistent with CO observations of this same region \citep{Salome2004,McDonald2012a}, which show that cold molecular hydrogen is present in the H$\alpha$ filaments but not to the west of the filaments.

Based on the physical arguments discussed in \cite{Fabian2001,McDonald2009}, the H$\alpha$ filament most likely arose from a ``cooling wake'', where X-ray bright ICM cools around the motion of the BCG. While the measured spectroscopic cooling rate of X-ray gas in the H$\alpha$ filament is consistent with this picture, the evidence for high speed motion of the cool core and the wider X-ray tail presented here may suggest a more complex picture where ram pressure stripping of the cool core (and not only the BCG) may be responsible for the full extent of the $\sim 3 \keV$ tail of ICM. This ICM tail is roughly the same width as the cool core itself, which is broadly consistent with expectations from the ram pressure stripping of a cool core. The obvious asymmetry across the eastern and western halves of this tail at other wavelengths, however, disfavor a scenario where the cool core is being uniformly stripped by ram pressure.

\subsection{The Cold Front}

 As stated above, the cooling wake scenario proposed by \cite{Fabian2001} to account for the H$\alpha$ filaments is strongly corroborated by these deeper observations of Abell 1795, which provide strong evidence for a cold front arising from the northward motion of the BCG. The large pressure jump across the cold front and the observed break in the surface brightness approximately 20 \kpc \ to the north of the cold front, however, suggest that the cool core may be traveling through the ICM at supersonic velocities. Further data will be necessary to conclusively resolve the presence of the bow shock predicted from this supersonic motion, however, as we cannot resolve the temperature and pressure structure of the ICM on sufficiently small scales to conclusively identify the presence of a shock.  

Despite the evidence from the cold front and surface brightness profile for supersonic motion, it is doubtful that the entire cool core is traversing at the these inferred speeds. The stripped tail of cool gas behind the BCG in Abell 1795 is only resolved out to distances of $\sim 50 \kpc$ despite the extremely deep exposure time and high spatial resolution of these \cha \ observations, which suggests that cool gas is likely absent at larger radii as opposed to being unresolved due to limited photon statistics or spatial resolution. The free-fall velocity ($\mysub{v}{ff}(r) = \sqrt{2 G M(<r)/r}$) of the cool core from a distance of $\sim 50 \kpc$ is calculated to be nearly a factor of two or three lower than the speed of the cool core as inferred from these observations \citep[using the mass profile for this cluster presented in][]{Vikhlinin2006}, which suggests that it is physically impossible for sloshing alone to be responsible for the cool core's measured speed. It is also likely that any merger capable of accelerating the cool core to these velocities is likely to destroy the cool core outright (ZuHone; private communication).} An infalling merger from larger distances may account for the necessary velocity, but such a scenario is disfavored by optical observations of this cluster. Usually in such a merger both the main cluster and the infalling subcluster initially host their own constituent galaxy population, including independent BCG's. There is no evidence for a second BCG in Abell 1795 in public {\it Hubble} or Sloan Digital Sky Survey images of Abell 1795, suggesting that no galaxy group scale (or larger) substructures are present in this cluster. Additionally, the galaxies initially hosted by the subcluster would continue to traverse northward through the main cluster ICM without being subject to the same drag forces as the cool core, separating from the cool core in the process. Since we observe no excess of galaxies to the north of the cool core, this also disfavors a scenario where the supersonic cool core is the remnant of an infalling galaxy group from beyond $\sim \rfive$. With these data we cannot explictly rule out a merger where the subcluster contains only diffuse gas with no galaxies, but given the necessary density of the cool gas such a merger would be disfavored on physical grounds. 
AGN feedback also likely plays a role in the extreme pressure jump across the cold front that we observe, as the radio jets are localized to precisely the region of this cold front. Careful modeling of AGN feedback in a cluster when the BCG is in motion with respect to the large scale cluster potential well will be required, however, to understand precisely how AGN feedback may locally accelerate the ICM in the cluster core. Further simulation work on cool core sloshing will also be necessary to confirm the expected velocities we expect for cool cores undergoing sloshing and the impact those motions may have on the core itself.

No matter what the cool core's bulk speed is, it is likely that this cool core will continue to be subject to major disruptions by the surrounding ICM. The final extent to which it will be diminished remains unclear, however. It is even possible that the bulk motion of the cool core may be sufficient to ultimately destroy it outright, although new hydrodynamic simulations that incorporate a more complete understanding of the astrophysics of the ICM will be required to better understand the nature of this merging galaxy cluster.

\section{Future Prospects}
 
Because Abell 1795 is the target of ongoing \cha \ calibration observations, deeper \cha \ data will become available over the telescope's remaining lifetime. These additional observations may allow for even higher resolution thermodynamic mapping and additional high precision analyses to be performed, including resolving the metallicity structure on smaller scales. In particular, the additional counts may provide the signal-to-noise needed to more conclusively the thermodynamic structure and nature of the candidate shock front to the north of the BCG. A companion study that utilizes these same data to probe the large scale thermodynamic structure of Abell 1795 is currently in preparation, which will allow us to connect these features observed at the center of the cluster to its larger scale thermodynamic structure.  

\section*{Acknowledgments}

We wish to thank John Zuhone for stimulating discussions regarding simulations of cool core sloshing, and the anonymous referee whose comments have greatly improved the manuscript. Support for this work (SE, MWB, EM) was provided by the Smithsonian Astrophysical Observatory (SAO) subcontract SV2-82023 under NASA contract NAS8-03060.  MM acknowledges support by NASA through a Hubble
Fellowship grant HST-HF51308.01-A awarded by the
Space Telescope Science Institute, which is operated by the
Association of Universities for Research in Astronomy, Inc.,
for NASA, under contract NAS 5-26555. This work was also supported
in part by the NASA grant GO8-9122X (LD). We also wish to thank the \cha \ X-ray Center's calibration, planning, operations, and data pipeline teams for acquiring the majority of these data. 

\bibliographystyle{apj}
\def \aap {A\&A} % alternative A&A code
\def \statisci {Statis. Sci.}
\def \physrep {Phys. Rep.}
\def \pre {Phys.\ Rev.\ E}
\def \sjos {Scand. J. Statis.} % Scandinavian Journal of Statistics
\def \jrssb {J. Roy. Statist. Soc. B} % Journal of the Royal Statistical Society. Series B (Statistical Methodology)

% note: one of the macros below causes the dvips output to go crazy on the references page
% uncomment as necessary, at your own risk

% This is ASTROMNEMONIC.BIB, a bibliography database file that provides 
% mnmonics for journal names, which can be used in the `journal' field of 
% bibliography entries in databases for BibTeX.
% The abbreviations of the most popular Astronomical journal names in 
% this file are set out in 1993 MNRAS 160, 1. For the other journals,
% the abbreviations of the journal names in this file follow the rules of
% the International List of Periodical Title Word Abbreviations.
% From: Astronomy and Astrophysics Abstracts, 1990, Vol. 49A
% As coded in mnemonic.bib, by Sake J. Hogeveen.
%
% H. Ferguson 19 May 1993

\def \araa {ARA\&A}
\def \aj {AJ}
 \def \aas {A\&AS}
  \def \aaps {A\&AS}
\def \apj {ApJ}
\def \apjl {ApJL}
\def \apjs {ApJS}
\def \mnras {MNRAS}
\def \nat {Nat}
 \def \pasj {PASJ}
 \def \pasp {PASP}
\def \gca {Geochim.\ Cosmochim.\ Acta}
\def \prd {Phys.\ Rev.\ D}
\def \prl {Phys.\ Rev.\ Lett.}

\bibliography{AllRefs}

\begin{thebibliography}{}
\expandafter\ifx\csname natexlab\endcsname\relax\def\natexlab#1{#1}\fi

\bibitem[{{Arnaud}(2004)}]{Arnaud2004}
{Arnaud}, K. 2004, in Bulletin of the American Astronomical Society, Vol.~36,
  Bulletin of the American Astronomical Society, 934--+

\bibitem[{{Ascasibar} \& {Markevitch}(2006)}]{Ascasibar2006}
{Ascasibar}, Y., \& {Markevitch}, M. 2006, \apj, 650, 102

\bibitem[{{Balucinska-Church} \& {McCammon}(1992)}]{McCammon1992}
{Balucinska-Church}, M., \& {McCammon}, D. 1992, \apj, 400, 699

\bibitem[{{Bautz} {et~al.}(2009){Bautz}, {Miller}, {Sanders}, {Arnaud},
  {Mushotzky}, {Porter}, {Hayashida}, {Henry}, {Hughes}, {Kawaharada},
  {Makashima}, {Sato}, \& {Tamura}}]{Bautz2009}
{Bautz}, M.~W., {Miller}, E.~D., {Sanders}, J.~S., {et~al.} 2009, \pasj, 61,
  1117

\bibitem[{{Blanton} {et~al.}(2011){Blanton}, {Randall}, {Clarke}, {Sarazin},
  {McNamara}, {Douglass}, \& {McDonald}}]{Blanton2011}
{Blanton}, E.~L., {Randall}, S.~W., {Clarke}, T.~E., {et~al.} 2011, \apj, 737,
  99

\bibitem[{{Cash}(1979)}]{Cash1979}
{Cash}, W. 1979, \apj, 228, 939

\bibitem[{{Cowie} {et~al.}(1983){Cowie}, {Hu}, {Jenkins}, \&
  {York}}]{Cowie1983}
{Cowie}, L.~L., {Hu}, E.~M., {Jenkins}, E.~B., \& {York}, D.~G. 1983, \apj,
  272, 29

\bibitem[{{Crawford} {et~al.}(2005{\natexlab{a}}){Crawford}, {Hatch}, {Fabian},
  \& {Sanders}}]{Crawford2005}
{Crawford}, C.~S., {Hatch}, N.~A., {Fabian}, A.~C., \& {Sanders}, J.~S.
  2005{\natexlab{a}}, \mnras, 363, 216

\bibitem[{{Crawford} {et~al.}(2005{\natexlab{b}}){Crawford}, {Sanders}, \&
  {Fabian}}]{Crawford2005b}
{Crawford}, C.~S., {Sanders}, J.~S., \& {Fabian}, A.~C. 2005{\natexlab{b}},
  \mnras, 361, 17

\bibitem[{{Ehlert} {et~al.}(2011){Ehlert}, {Allen}, {von der Linden},
  {Simionescu}, {Werner}, {Taylor}, {Gentile}, {Ebeling}, {Allen}, {Applegate},
  {Dunn}, {Fabian}, {Kelly}, {Million}, {Morris}, {Sanders}, \&
  {Schmidt}}]{Ehlert2011}
{Ehlert}, S., {Allen}, S.~W., {von der Linden}, A., {et~al.} 2011, \mnras, 411,
  1641

\bibitem[{{Ettori} {et~al.}(2002){Ettori}, {Fabian}, {Allen}, \&
  {Johnstone}}]{Ettori2002}
{Ettori}, S., {Fabian}, A.~C., {Allen}, S.~W., \& {Johnstone}, R.~M. 2002,
  \mnras, 331, 635

\bibitem[{{Fabian} {et~al.}(2008){Fabian}, {Johnstone}, {Sanders}, {Conselice},
  {Crawford}, {Gallagher}, \& {Zweibel}}]{Fabian2008}
{Fabian}, A.~C., {Johnstone}, R.~M., {Sanders}, J.~S., {et~al.} 2008, \nat,
  454, 968

\bibitem[{{Fabian} {et~al.}(2001){Fabian}, {Sanders}, {Ettori}, {Taylor},
  {Allen}, {Crawford}, {Iwasawa}, \& {Johnstone}}]{Fabian2001}
{Fabian}, A.~C., {Sanders}, J.~S., {Ettori}, S., {et~al.} 2001, \mnras, 321,
  L33

\bibitem[{{Fabian} {et~al.}(2011){Fabian}, {Sanders}, {Williams}, {Lazarian},
  {Ferland}, \& {Johnstone}}]{Fabian2011}
{Fabian}, A.~C., {Sanders}, J.~S., {Williams}, R.~J.~R., {et~al.} 2011, \mnras,
  417, 172

\bibitem[{{Forman} {et~al.}(2005){Forman}, {Nulsen}, {Heinz}, {Owen}, {Eilek},
  {Vikhlinin}, {Markevitch}, {Kraft}, {Churazov}, \& {Jones}}]{Forman2005}
{Forman}, W., {Nulsen}, P., {Heinz}, S., {et~al.} 2005, \apj, 635, 894

\bibitem[{{Ge} \& {Owen}(1993)}]{Ge1993}
{Ge}, J.~P., \& {Owen}, F.~N. 1993, \aj, 105, 778

\bibitem[{{Henry} {et~al.}(2004){Henry}, {Finoguenov}, \& {Briel}}]{Henry2004}
{Henry}, J.~P., {Finoguenov}, A., \& {Briel}, U.~G. 2004, \apj, 615, 181

\bibitem[{{Jaffe} {et~al.}(2005){Jaffe}, {Bremer}, \& {Baker}}]{Jaffe2005}
{Jaffe}, W., {Bremer}, M.~N., \& {Baker}, K. 2005, \mnras, 360, 748

\bibitem[{{Kaastra} \& {Mewe}(1993)}]{Kaastra1993}
{Kaastra}, J.~S., \& {Mewe}, R. 1993, \aas, 97, 443

\bibitem[{{Kalberla} {et~al.}(2005){Kalberla}, {Burton}, {Hartmann}, {Arnal},
  {Bajaja}, {Morras}, \& {P{\"o}ppel}}]{Kalberla2005}
{Kalberla}, P.~M.~W., {Burton}, W.~B., {Hartmann}, D., {et~al.} 2005, \aap,
  440, 775

\bibitem[{{Landau} \& {Lifshitz}(1959)}]{Landau1959}
{Landau}, L.~D., \& {Lifshitz}, E.~M. 1959, {Fluid mechanics}

\bibitem[{{Liedahl} {et~al.}(1995){Liedahl}, {Osterheld}, \&
  {Goldstein}}]{Liedhal1995}
{Liedahl}, D.~A., {Osterheld}, A.~L., \& {Goldstein}, W.~H. 1995, \apjl, 438,
  L115

\bibitem[{{Lodders}(2003)}]{Lodders2003}
{Lodders}, K. 2003, \apj, 591, 1220

\bibitem[{{Mahdavi} {et~al.}(2005){Mahdavi}, {Finoguenov}, {B{\"o}hringer},
  {Geller}, \& {Henry}}]{Mahdavi2005}
{Mahdavi}, A., {Finoguenov}, A., {B{\"o}hringer}, H., {Geller}, M.~J., \&
  {Henry}, J.~P. 2005, \apj, 622, 187

\bibitem[{{Maloney} \& {Bland-Hawthorn}(2001)}]{Maloney2001}
{Maloney}, P.~R., \& {Bland-Hawthorn}, J. 2001, \apjl, 553, L129

\bibitem[{{Markevitch} \& {Vikhlinin}(2007)}]{Markevitch2007}
{Markevitch}, M., \& {Vikhlinin}, A. 2007, \physrep, 443, 1

\bibitem[{{McDonald} \& {Veilleux}(2009)}]{McDonald2009}
{McDonald}, M., \& {Veilleux}, S. 2009, \apjl, 703, L172

\bibitem[{{McDonald} {et~al.}(2011){McDonald}, {Veilleux}, \&
  {Mushotzky}}]{McDonald2011a}
{McDonald}, M., {Veilleux}, S., \& {Mushotzky}, R. 2011, \apj, 731, 33

\bibitem[{{McDonald} {et~al.}(2012{\natexlab{a}}){McDonald}, {Veilleux}, \&
  {Rupke}}]{McDonald2012b}
{McDonald}, M., {Veilleux}, S., \& {Rupke}, D.~S.~N. 2012{\natexlab{a}}, \apj,
  746, 153

\bibitem[{{McDonald} {et~al.}(2010){McDonald}, {Veilleux}, {Rupke}, \&
  {Mushotzky}}]{McDonald2010}
{McDonald}, M., {Veilleux}, S., {Rupke}, D.~S.~N., \& {Mushotzky}, R. 2010,
  \apj, 721, 1262

\bibitem[{{McDonald} {et~al.}(2012{\natexlab{b}}){McDonald}, {Wei}, \&
  {Veilleux}}]{McDonald2012a}
{McDonald}, M., {Wei}, L.~H., \& {Veilleux}, S. 2012{\natexlab{b}}, \apjl, 755,
  L24

\bibitem[{{McNamara} {et~al.}(1996){McNamara}, {Jannuzi}, {Elston}, {Sarazin},
  \& {Wise}}]{McNamara1996}
{McNamara}, B.~R., {Jannuzi}, B.~T., {Elston}, R., {Sarazin}, C.~L., \& {Wise},
  M. 1996, \apj, 469, 66

\bibitem[{{McNamara} \& {Nulsen}(2007)}]{McNamara2007}
{McNamara}, B.~R., \& {Nulsen}, P.~E.~J. 2007, \araa, 45, 117

\bibitem[{{McNamara} \& {Nulsen}(2012)}]{McNamara2012}
---. 2012, New Journal of Physics, 14, 055023

\bibitem[{{Million} {et~al.}(2010){Million}, {Allen}, {Werner}, \&
  {Taylor}}]{Million2010}
{Million}, E.~T., {Allen}, S.~W., {Werner}, N., \& {Taylor}, G.~B. 2010,
  \mnras, 582

\bibitem[{{Oegerle} \& {Hill}(2001)}]{Oegerle2001}
{Oegerle}, W.~R., \& {Hill}, J.~M. 2001, \aj, 122, 2858

\bibitem[{{Salom{\'e}} \& {Combes}(2004)}]{Salome2004}
{Salom{\'e}}, P., \& {Combes}, F. 2004, \aap, 415, L1

\bibitem[{{Sanders}(2006)}]{Sanders2006}
{Sanders}, J.~S. 2006, \mnras, 371, 829

\bibitem[{{Sanders} \& {Fabian}(2007)}]{Sanders2007}
{Sanders}, J.~S., \& {Fabian}, A.~C. 2007, \mnras, 381, 1381

\bibitem[{{Simionescu} {et~al.}(2009){Simionescu}, {Werner}, {B{\"o}hringer},
  {Kaastra}, {Finoguenov}, {Br{\"u}ggen}, \& {Nulsen}}]{Simionescu2009}
{Simionescu}, A., {Werner}, N., {B{\"o}hringer}, H., {et~al.} 2009, \aap, 493,
  409

\bibitem[{{Sun} {et~al.}(2007){Sun}, {Donahue}, \& {Voit}}]{Sun2007}
{Sun}, M., {Donahue}, M., \& {Voit}, G.~M. 2007, \apj, 671, 190

\bibitem[{{van Breugel} {et~al.}(1984){van Breugel}, {Heckman}, \&
  {Miley}}]{VanBreugel1984}
{van Breugel}, W., {Heckman}, T., \& {Miley}, G. 1984, \apj, 276, 79

\bibitem[{{Vikhlinin} {et~al.}(2006){Vikhlinin}, {Kravtsov}, {Forman}, {Jones},
  {Markevitch}, {Murray}, \& {Van Speybroeck}}]{Vikhlinin2006}
{Vikhlinin}, A., {Kravtsov}, A., {Forman}, W., {et~al.} 2006, \apj, 640, 691

\bibitem[{{Werner} {et~al.}(2010){Werner}, {Simionescu}, {Million}, {Allen},
  {Nulsen}, {von der Linden}, {Hansen}, {B{\"o}hringer}, {Churazov}, {Fabian},
  {Forman}, {Jones}, {Sanders}, \& {Taylor}}]{Werner2010}
{Werner}, N., {Simionescu}, A., {Million}, E.~T., {et~al.} 2010, \mnras, 407,
  2063

\bibitem[{{Werner} {et~al.}(2012){Werner}, {Oonk}, {Canning}, {Allen},
  {Simionescu}, {Kos}, {van Weeren}, {Edge}, {Fabian}, {von der Linden},
  {Nulsen}, {Reynolds}, \& {Ruszkowski}}]{Werner2012}
{Werner}, N., {Oonk}, J.~B.~R., {Canning}, R.~E.~A., {et~al.} 2012, \apj \
  submitted [arXive:1211.6722], arXiv:1211.6722

\bibitem[{{ZuHone} {et~al.}(2013){ZuHone}, {Markevitch}, {Brunetti}, \&
  {Giacintucci}}]{ZuHone2013}
{ZuHone}, J.~A., {Markevitch}, M., {Brunetti}, G., \& {Giacintucci}, S. 2013,
  \apj, 762, 78

\bibitem[{{ZuHone} {et~al.}(2010){ZuHone}, {Markevitch}, \&
  {Johnson}}]{ZuHone2009}
{ZuHone}, J.~A., {Markevitch}, M., \& {Johnson}, R.~E. 2010, \apj, 717, 908

\end{thebibliography}

\end{document}